\documentclass[aps,pra,twocolumn,showpacs,superscriptaddress,floatfix,10pt]{revtex4-1}
\usepackage{framed}
\usepackage{graphicx}
\usepackage{amsmath}
\usepackage{epsfig}
\usepackage{helvet}
\usepackage{amssymb}
\usepackage{framed}
\usepackage{hyperref}

\begin{document}

\title{A tensor network quotient takes the vacuum to the thermal state\\
}

\author{Bart{\l}omiej Czech} 
\affiliation{Department of Physics, Stanford University, Stanford, CA 94305-4045}
\author{Glen Evenbly} 
\affiliation{Department of Physics and Astronomy, University of California, Irvine, CA 92697-4575}
\author{Lampros Lamprou} 
\affiliation{Department of Physics, Stanford University, Stanford, CA 94305-4045}
\author{\mbox{Samuel McCandlish}} 
\affiliation{Department of Physics, Stanford University, Stanford, CA 94305-4045}
\author{Xiao-liang Qi} 
\affiliation{Department of Physics, Stanford University, Stanford, CA 94305-4045}
\author{James Sully}
\affiliation{SLAC National Accelerator Laboratory, Menlo Park, CA 94025}
\author{Guifr{\'e} Vidal}
\affiliation{Perimeter Institute for Theoretical Physics, Waterloo, Ontario N2L 2Y5, Canada}
\vskip 0.25cm
\date{\today}

\begin{abstract}
\noindent
In 1+1-dimensional conformal field theory, the thermal state on a circle is related to a certain quotient of the vacuum on a line. We explain how to take this quotient in the MERA tensor network representation of the vacuum and confirm the validity of the construction in the critical Ising model. This result suggests that the tensors comprising MERA can be interpreted as performing local scale transformations, so that adding or removing them emulates conformal maps. In this sense, the optimized MERA recovers local conformal invariance, which is explicitly broken by the choice of lattice. Our discussion also informs the dialogue between tensor networks and holographic duality.
\end{abstract}


\maketitle

Tensor networks provide an efficient method for describing wave-functions of complex quantum systems. A key question that arises in their application concerns the symmetries preserved by the network. In this letter, we consider the Multi-scale Entanglement Renormalization Ansatz (MERA) \cite{mera}, a special network with properties attuned to systems at criticality. In the continuum, such systems are typically described by conformal field theories (CFT) and the relevant symmetry is the global conformal group. In 1+1 dimensions, however, the conformal symmetry is additionally enhanced by \emph{local} conformal transformations \cite{BPZ}. Na{\"\i}vely, any tensor network realization of the ground state wave-function automatically breaks this symmetry through a choice of discretization. The primary goal of the present paper is to explain that, in fact, the optimized MERA network restores a discrete form of local conformal invariance.

Local conformal symmetry has an interesting consequence, which we use as a test of local conformal invariance. The Euclidean path integrals that prepare the thermal state on a circle and the vacuum on an infinite line are related through (i) a local conformal map and (ii) a subsequent quotient by a discrete scale transformation. If a tensor network representing the ground state manifests local conformal symmetry, it ought to give rise to a similar relation. We shall see that the ground state MERA allows us to perform the required local conformal map and obtain a network that is manifestly invariant under discrete scale transformations. We will then implement the quotient and, in the case of the critical Ising model, confirm that the resulting network correctly prepares the thermal state. This provides strong evidence that the optimized ground state MERA has an emergent local scale invariance.

Our construction has further consequences, which we discuss in the closing section. The quotient network naturally decomposes into three types of regions: an isometry, an approximate isometry, and a region responsible for the spectrum of the state. The three are distinguished with reference to the auxiliary causal structure in the MERA network, with the quotient symmetry acting in a spacelike, lightlike and timelike fashion, respectively. This decomposition is pertinent for the debate about the relation between tensor networks and the AdS/CFT correspondence \cite{adscft}, which was initiated in \cite{swingle}. We sketch those consequences only briefly, referring the holographically inclined reader to a future, more detailed publication \cite{secondpaper}.

\begin{figure}[!t]
\begin{center}
\includegraphics[width=8.5cm]{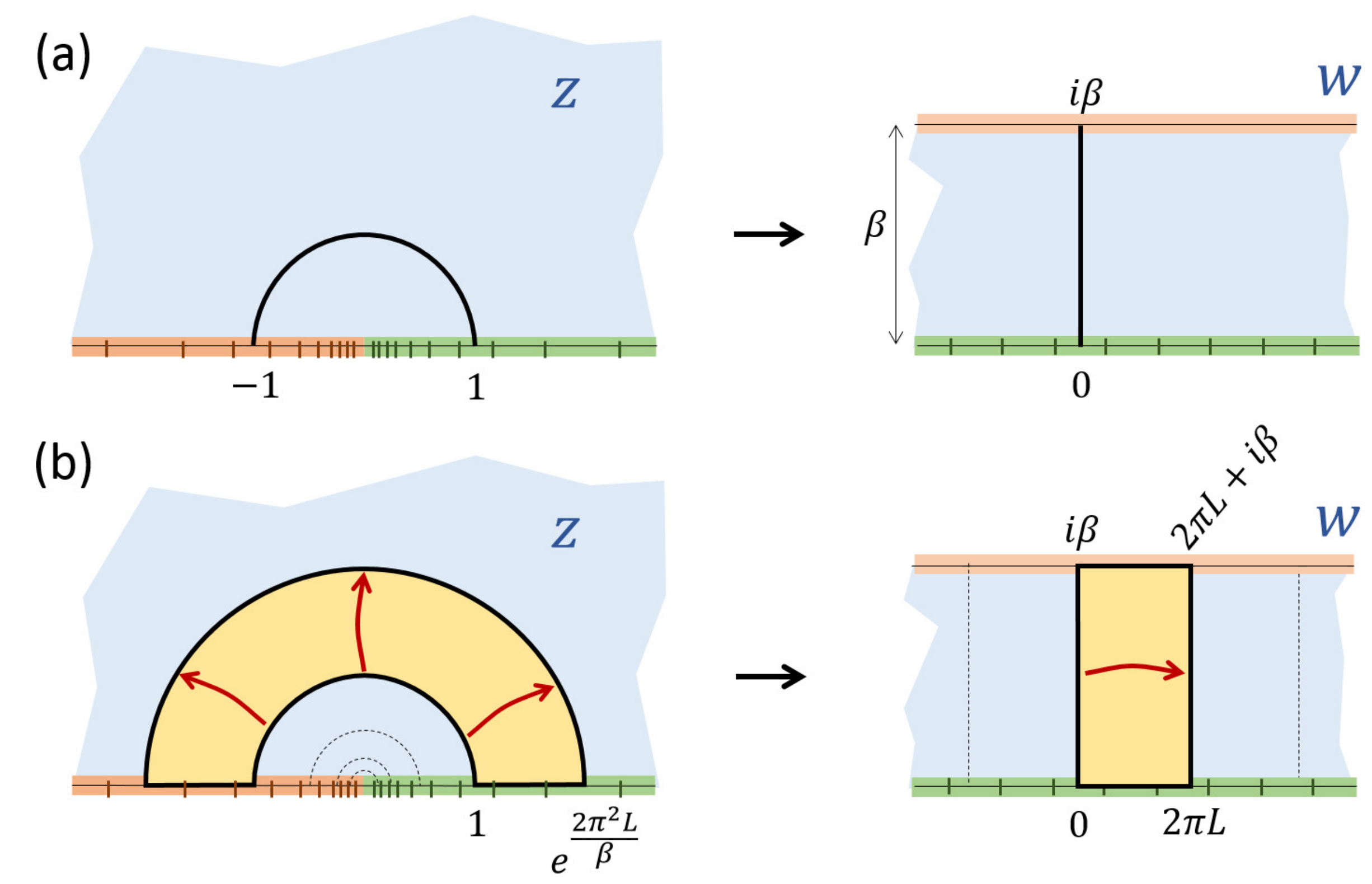}
\caption{
(a) The conformal map $z \rightarrow w = (\beta/\pi)\log z$ maps the upper half plane to an infinite strip of width $\beta$. An exponentially spaced discretization of the positive real axis on the $z$ plane is mapped to a regular discretization of the real axis on the $w$ plane. 
(b) A quotient by a scaling transformation by a factor $\exp (2\pi^2L/\beta)$ in the upper half plane produces a topological cylinder.
That scaling amounts to a translation by $2\pi L$ in the infinite strip. A quotient by a translation by $2\pi L$ in the infinite strip produces a flat cylinder with tallness $\beta$ and circumference of the base equal to $2\pi L$.  
}
\label{fig:zw}
\end{center}
\end{figure}

\textit{Quotient in the path integral.---} Consider the Euclidean path integral for a CFT$_2$ on the upper half-plane. This object prepares the vacuum state on an infinite line. A conformal transformation $z\rightarrow w =  (\beta/\pi)\log z$ maps the upper half plane to an infinite flat strip of width $\beta$. By standard field theory arguments, the Euclidean path integral on this strip prepares the thermal state on an infinite line with temperature $T=\frac{1}{\beta}$, see fig.~\ref{fig:zw}(a).

Now, exploiting the translational invariance of the new path integral, we quotient the infinite strip by a discrete translation $w\sim w+2\pi L$. This amounts to identifying two vertical lines separated by $2\pi L$ to produce a flat cylinder with height $\beta$ and radius $L$. The Euclidean path integral on this cylinder prepares the thermal state on a circle of radius $L$ with inverse temperature $\beta$. It is instructive to understand the action of the quotient in terms of the upper half-plane: The two identified vertical lines in the infinite strip come from a pair of concentric semi-circles in the upper half-plane, with radii related by a relative factor of $\exp(2\pi^2L/\beta)$, see fig.~\ref{fig:zw}(b). The thermal state on the circle can, thus, be obtained by quotienting the original path integral by a discrete scale transformation, producing a topological cylinder which is then flattened by an application of the $\log z$ map.

It is convenient to introduce the reduced inverse temperature $\tilde{\beta} \equiv \beta/(2\pi L)$, which is invariant under global scalings of the cylinder. In a conformal field theory, the spectrum of  the thermal density operator depends only on $\tilde{\beta}$ and reads:
\begin{eqnarray} \label{eq:spectrum}
\lambda_{\alpha} &=& \frac{1}{Z(\tilde{\beta})} e^{-2\pi \tilde{\beta} \left(\Delta_{\alpha} - \frac{c}{12}\right)} \\ 
Z(\tilde{\beta}) &\equiv& \sum_{\alpha} e^{-2\pi\tilde{\beta} \left(\Delta_{\alpha} - \frac{c}{12} \right)}
\end{eqnarray}
Here $c$ is the central charge, $\Delta_{\alpha}$ are the scaling dimensions of local operators, and $Z(\tilde{\beta})$ stands for the partition function of the CFT, which is computed by the path integral on a torus with imaginary modular parameter $\tau = i\tilde{\beta}$. The infinite strip of fig.~\ref{fig:zw}(a) can, therefore, be regarded as the $L \rightarrow \infty$ limit of the cylinder and it prepares a thermal state with vanishing reduced inverse temperature, $\tilde{\beta}=0$.

In summary, the Euclidean path integral preparing the CFT ground state on an infinite line can be transformed to a representation of the thermal state on a circle by (i) a local conformal map from the upper half-plane to an infinite strip of width $\beta$ and (ii) a subsequent quotient by a discrete translation, which on the initial half-plane acts as a discrete rescaling. As we explain below, operations (i) and (ii) have analogues in the discrete setting of MERA.

\begin{figure}[!t]
\begin{center}
\includegraphics[width=8.5cm]{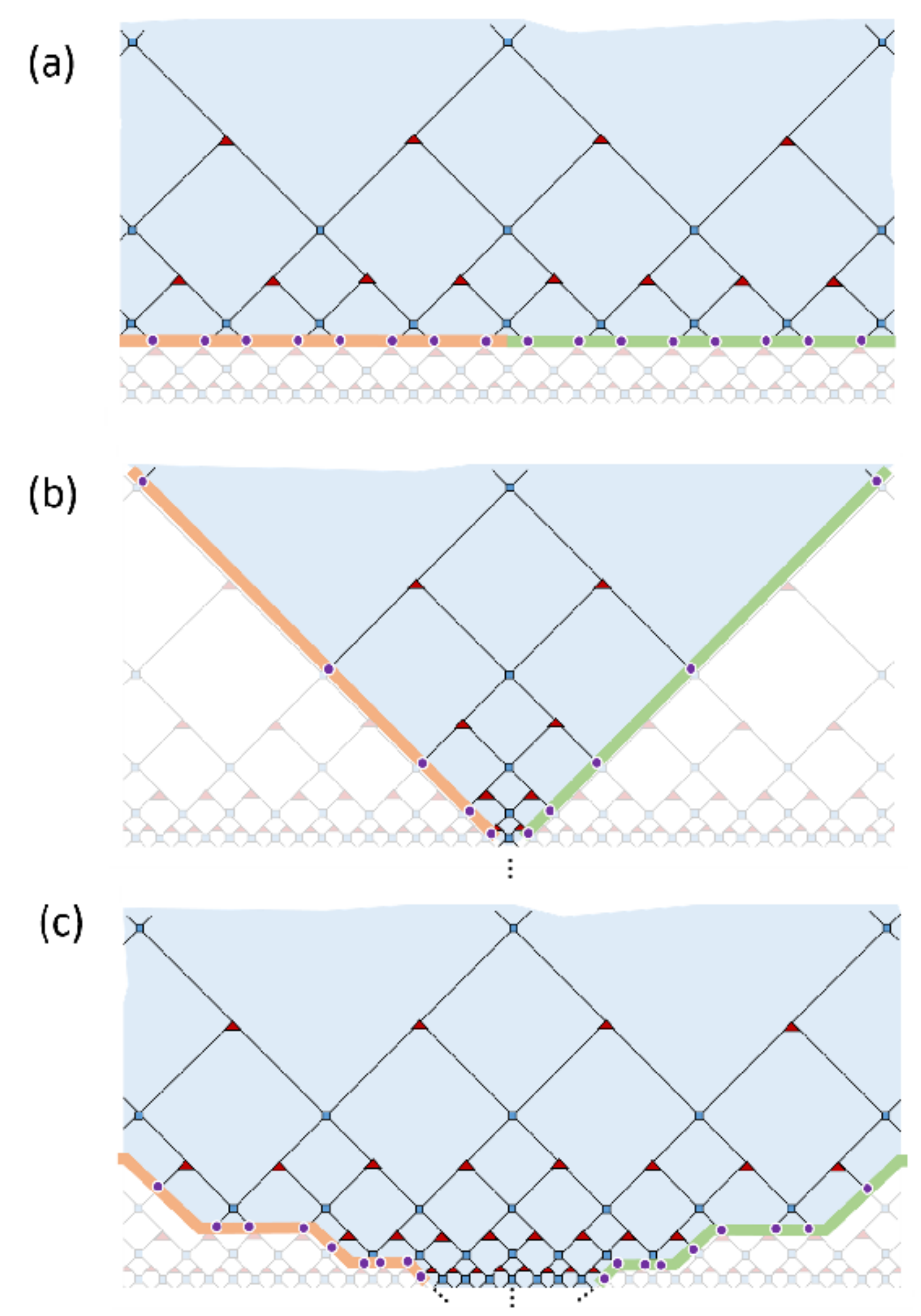}
\caption{
(a) MERA for the ground state on a uniformly discretized infinite line.
(b) We can use disentanglers and isometries to replace the regular discretization with an exponential discretization. This produces infinitely many sites near the origin, effectively disconnecting the positive and negative parts of the real axis. This local coarse-graining is a lattice version of the logarithmic scaling map (\ref{log2}).
(c) A logarithmic scaling map with a different prefactor.
}
\label{fig:local}
\end{center}
\end{figure}

\textit{From the Euclidean path integral to MERA.---} Generally, we expect facts about the Euclidean path integral to be realizable in the MERA tensor network, because the former can be transformed into the latter by Tensor Network Renormalization (TNR) \cite{tnr, tnr2mera}. The TNR algorithm involves two main ingredients: a choice of discretization of the spatial axis and iterative coarse-grainings of the appropriately discretized path integral. For the path integral on a half-plane, choosing a uniform discretization on the real line leads to the MERA tensor network shown in fig. \ref{fig:local}(a). 

However, the selection of a particular discretization breaks conformal invariance. Indeed, a local rescaling transforms distances on the spatial axis according to $z \to f(z)$, which can map any lattice to any other one. (An example central to this paper is the logarithmic map of fig.~\ref{fig:zw}, which takes an exponentially spaced set of cuts to the uniform one.) Thus, the route from the Euclidean path integral to MERA requires breaking conformal invariance. Our interest is in showing that the optimized MERA restores a discrete form of this symmetry.

\textit{Local scale transformations on the lattice.---} MERA is a network of tensors with a set pattern of contractions, numerically optimized to describe the ground state of a CFT on a 1-dimensional lattice, see fig.~\ref{fig:local}(a). The two types of tensors used, the disentanglers and the isometries, acquire the roles of removing short-range entanglement and coarse-graining the state, respectively. If we remove one layer of the network---a row of disentanglers and a row of isometries---we effect a coarse-graining by a constant factor $\lambda = 1/2$. We may also go in the opposite direction and fine-grain the lattice by adding an extra layer. Each additional row of disentanglers and isometries performs a global change of scale by a relative factor $\lambda = 2$ (see Appendix~A for details).

But disentanglers and isometries need not be applied globally, as an entire row. We may also apply individual disentanglers and isometries on parts of the network, leaving the rest of it intact. We conjecture that doing so effects a local scale transformation. In particular, erasing or adjoining tensors allows us to change the discretization of the line in inhomogeneous ways. 

As an example, figs.~\ref{fig:local}(b)-(c) show two choices of exponential discretizations. Measuring spatial distances with respect to these discretizations amounts to applying a logarithmic transformation. Specifically, going from fig.~\ref{fig:local}(a) to fig.~\ref{fig:local}(b) is a lattice version of the logarithmic map:
\begin{equation}
z \to w = \log_2 z = \frac{1}{\pi}\frac{\pi}{\log 2}\log z
\label{log2}
\end{equation}
This map is enacted by erasing and adjoining certain tensors in fig.~\ref{fig:local}(a) to obtain fig.~\ref{fig:local}(b). Likewise, erasing and adjoining tensors to take fig.~\ref{fig:local}(a) to fig.~\ref{fig:local}(c) also emulates a logarithmic map, though with a different base: 
\begin{equation}
z \to \tilde{w} = \log_{2^{1/4}} z = 4 \cdot \frac{1}{\pi}\frac{\pi}{\log 2}\log z
\label{4log2}
\end{equation}
Of course, this statement is subject to usual artifacts of discretization: for instance, the open indices in fig.~\ref{fig:local}(c) are not evenly spaced on the cut. Ignoring this subtlety, we observe that maps (\ref{log2}) and (\ref{4log2}) differ by a global change of scale by $\lambda = 4$. We can see this in the network by comparing figs.~\ref{fig:local}(b) and (c): tensors included in one but not the other form a uniform row of basic units, each of which takes four lines into one. (This is highlighted in fig.~\ref{fig:explicit} in Appendix~B).

We have seen earlier that the logarithmic map $z \to w = (\beta/\pi) \log z$ takes the path integral for the vacuum to one that prepares the thermal state. Transformations (\ref{log2}) and (\ref{4log2}) are of this form, with inverse temperatures $\beta = \pi/\log 2$ and $4\pi / \log2$. Thus, associating MERA tensors with local scale transformations predicts that the states prepared in figs.~\ref{fig:local}(b) and (c) will have thermal spectra. Verifying this prediction directly is computationally intractable. In order to circumvent this problem, we will take a quotient by a discrete scaling symmetry. The prediction is that the quotient of the state in fig.~\ref{fig:local}(b) (respectively (c)) will prepare the thermal state on a circle with inverse temperature $\beta = \pi/\log 2$ (respectively $4\pi/\log 2$).

\begin{figure}[!t]
\begin{center}
\includegraphics[width=8.5cm]{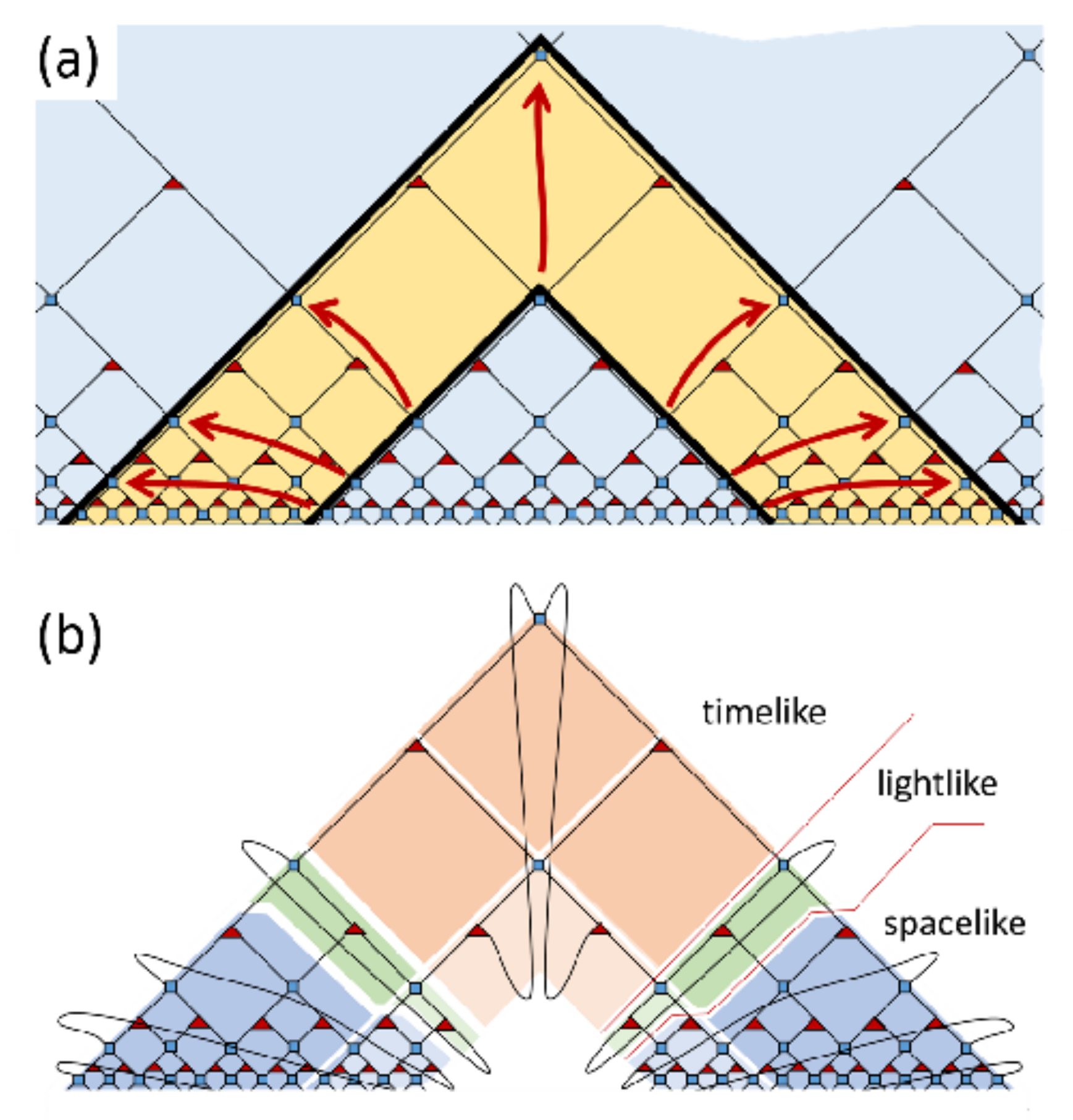}
\caption{
(a) A global scaling by $\lambda = 1/2$ identifies tensors in MERA. This becomes an exact symmetry of the network after applying a logarithmic scaling map, see figs.~\ref{fig:local}(b)-(c).
(b) The quotient of MERA by a scale factor $\lambda = (1/2)^k$ for $k=2$.
}
\label{fig:quotient}
\end{center}
\end{figure}

\textit{A MERA quotient for the thermal state on a circle.---} After applying the logarithmic maps, the resulting networks (figs.~\ref{fig:local}(b) and (c)) develop a scaling symmetry. This symmetry, shown in fig.~\ref{fig:quotient}(a), changes scales by a factor of $\lambda = 1/2$. Of course, this symmetry is present in the path integral in the continuum, but it was explicitly broken by the choice of discretization. Our discrete logarithmic maps restore it.

On the continuum $w$-strip, the scaling symmetry is represented as a translation, and quotienting by it yields a theory on a spatial circle. Now that our logarithmic maps have restored this symmetry in the tensor network, we are ready to take the analogous quotient. This produces a tensor network representation of the thermal state on a spatial circle. The quotient is implemented by a contraction of indices shown in fig.~\ref{fig:quotient}(b). The discrete nature of the network only allows quotienting by powers of $1/2$. We call the relevant exponent $k$ and consider the network obtained from modding out discrete $(1/2)^k$ scalings.

Depending on whether we quotient the network shown in fig.~\ref{fig:local}(b) or (c), the size of the spatial circle will differ by a factor of 4. More generally, we could consider applying the quotient to a network obtained by a discrete version of the logarithmic map 
\begin{equation}
z \to \tilde{w} = \log_{2^{1/p}} z = p \cdot \frac{1}{\pi}\frac{\pi}{\log 2}\log z\,,
\label{plog2}
\end{equation}
with the size of the spatial circle becoming $2\pi L = k\, p$. Notice, however, that the inverse temperature $\beta$ is also proportional to $p$: comparing eq.~(\ref{plog2}) with $w = (\beta / \pi) \log z$ gives $\beta = p\, \pi / \log 2$. This means that the choice of the logarithmic map does not affect the spectrum of the state, because the latter (viz. eq.~\ref{eq:spectrum}) depends only on the \emph{reduced} inverse temperature:
\begin{equation}
\tilde{\beta} = \frac{\beta}{2\pi L} = \frac{p\, \pi / \log 2}{k\, p} = \frac{\pi}{k \log 2}
\label{redbeta}
\end{equation}
Physically, $p$ drops out from $\tilde\beta$ because its effect is a global rescaling of the Euclidean cylinder on which the thermal state is prepared. 

That the spectrum of the quotient is independent of the overall scale of the logarithmic map highlights the importance of the causal structure of MERA reviewed in Appendix~A. After the quotient, the network splits into three types of regions, where the identification acts in a spacelike, lightlike, and timelike manner; see fig.~\ref{fig:quotient}(b). As we show in Appendix~B by a direct manipulation of tensors, the spacelike-identified component is an exact isometry. The lightlike-identified component is an approximate isometry when $k$ is not too small; we discuss this important point in more detail below. Because they are isometries, these regions do not affect the spectrum of the state; their role is instead to perform a change of basis. This change of basis is precisely the global coarse-graining that relates different logarithmic maps. We saw an example of this when we compared the networks in figs.~\ref{fig:local}(b) and (c): the tensors that distinguish them make up the lightlike-identified and the top layer of the spacelike-identified component of the quotient.

The isometric character of the spacelike-identified regions implies that the spectrum of the quotient network is prepared entirely in the lightlike and timelike-identified sectors. We have already mentioned that the lightlike-identified regions are approximately isometric, so for practical purposes they too drop out from determining the spectrum. If, as we conjecture, erasing and adjoining tensors in the optimized MERA performs local changes of scale, the state prepared in the timelike-identified piece of fig.~\ref{fig:quotient}(b) has the thermal spectrum with reduced inverse temperature $\tilde{\beta} = \pi / k \log 2$.

\begin{figure}[!t]
\begin{center}
\includegraphics[trim = 32mm 73mm 28mm 72mm, clip, width=8.5cm]{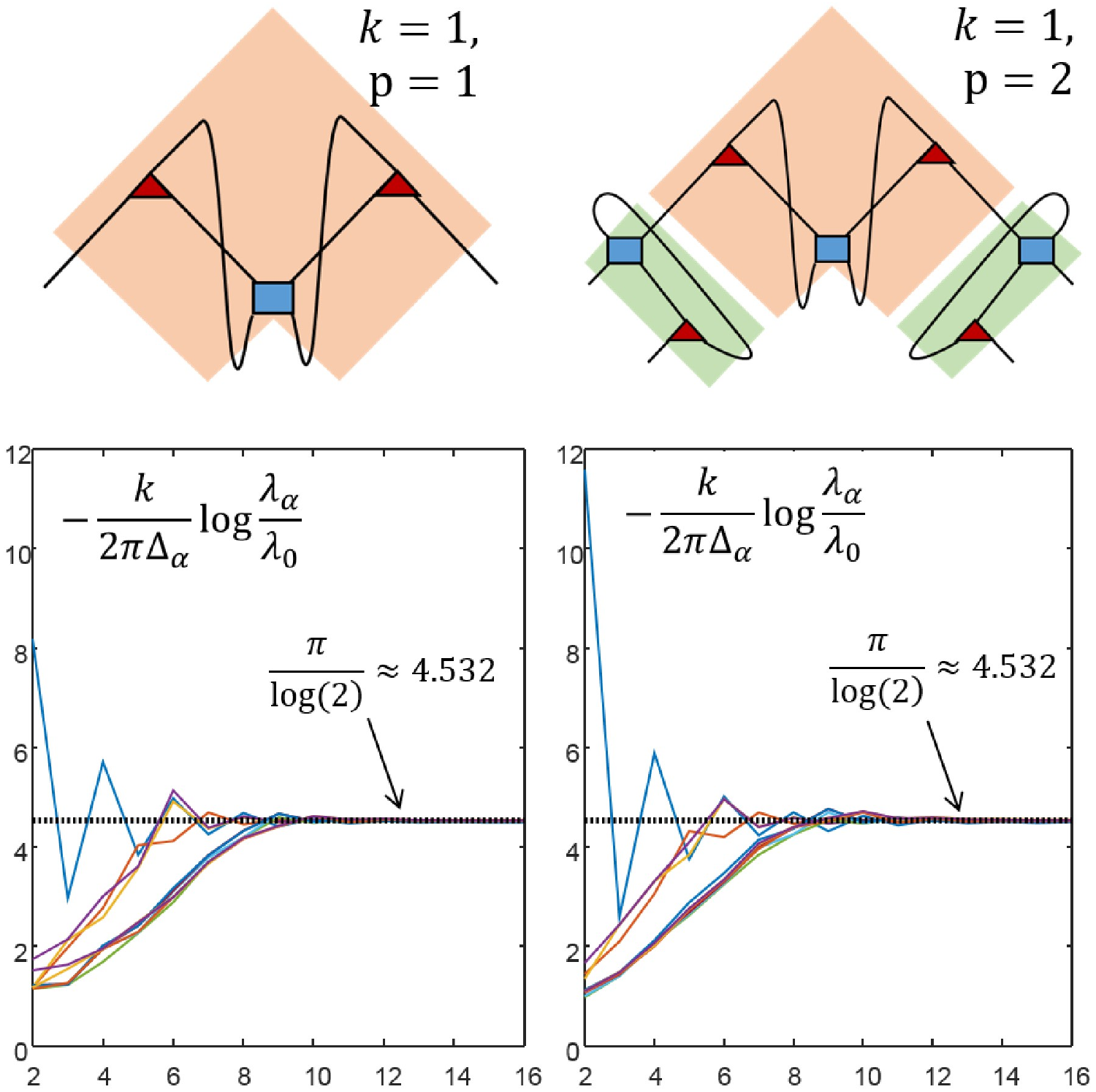}
\caption{
(Left) Spectrum of singular values $\lambda_{\alpha}$ of the quotient network (understood as a matrix between left and right open indices) for $k=1,2, \cdots, 16$ in the critical Ising model. The spectrum was obtained from the timelike-identified part of the quotient. For each singular value $\lambda_{\alpha}$, we plot the expression $-(k/2\pi\Delta_{\alpha})\log(\lambda_{\alpha}/\lambda_0)$, where $\Delta_{\alpha}$ are the exact scaling dimensions of the Ising CFT. In the absence of approximations, eqs.~(\ref{eq:spectrum}) and (\ref{redbeta}) would set this quantity to $\pi/\log 2$. (Right) The spectrum after including the nearly isometric, lightlike-identified regions in the quotient is almost identical except for small values of $k$.
}
\label{fig:numerics}
\end{center}
\end{figure}

\textit{Example: quantum Ising model.---} As a concrete example, we checked this claim in the 1+1-dimensional quantum Ising spin chain at criticality. Starting from an optimized MERA approximation (with bond dimension $\chi=4$) of the ground state on the infinite line obtained with TNR \cite{tnr, tnrAlgorithms}, we have built the quotient for $k \in [1, \cdots 16]$. Fig.~\ref{fig:numerics} shows that the spectrum of the quotient indeed quickly converges to the thermal spectrum of eq.~(\ref{eq:spectrum}) with the predicted value of $\tilde{\beta}$. The figure also shows that for large $k$ the lightlike-identified part of the quotient does not modify the spectrum. This indicates that this part of the network is an approximate isometry at sufficiently large $k$ (high enough temperature).

\textit{Discussion.---} The tensor network manipulations conducted in this paper illustrate how the optimized MERA encodes local conformal symmetry. A key feature of the vacuum MERA is that its open indices on the spatial axis are uniformly spaced; see fig.~\ref{fig:local}(a). Our construction of the thermal state called for an exponentially spaced set of indices such as those shown in the left of fig.~\ref{fig:zw}. To identify such a set, we erased and added some tensors, obtaining the networks shown in figs.~\ref{fig:local}(b) and (c). If we declare the open indices on these new networks to be uniformly spaced, the procedure of erasing or adding tensors acquires the interpretation of a local conformal transformation:
\begin{equation}
z \to w = (\beta / \pi) \log z
\end{equation} 
The fact that a subsequent quotient of the network reproduces the  thermal state on a circle confirms that our manipulations correctly emulated the logarithmic map. 

\begin{figure}[!t]
\begin{center}
\includegraphics[width=8.5cm]{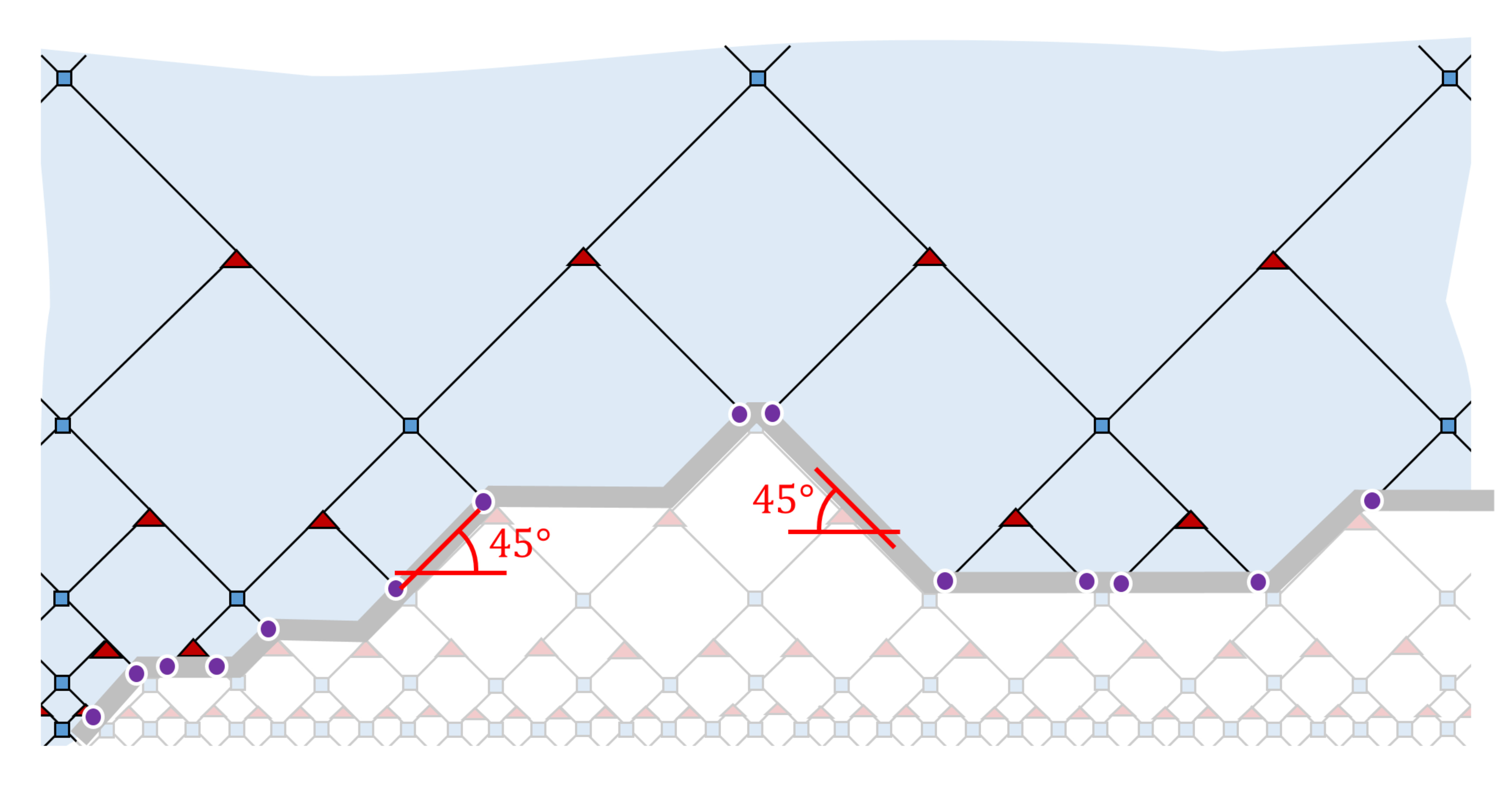}
\caption{
Adding or removing disentanglers and isometries locally implements a discrete local scale transformation. The open indices after the transformation live on a cut through MERA, which is piecewise horizontal (spacelike) or at 45$^\circ$ (lightlike) but never timelike.
}
\label{fig:local2}
\end{center}
\end{figure}

To implement a more general map $z \to w = f(z)$, we can imagine proceeding in a similar fashion. The idea is to look for a set of indices with spatial locations that approximate $z = f^{-1}(n)$ for $n \in \mathbb{Z}$. This encodes the function $f(z)$ in terms of a meandering cut through MERA such as the one shown in fig.~\ref{fig:local2}. The height at which the cut is drawn reflects the step between consecutive ticks, $f^{-1}(n+1)$ and $f^{-1}(n)$, which is how $f(z)$ resets a local scale. In terms of the causal structure of MERA, only piecewise spacelike and lightlike cuts are sensible, because timelike cuts are not related to the regular lattice by an isometry (see also Appendix~A). Of course, the discrete nature of the tensor network restricts the class of functions $f(z)$ that can be adequately represented by such a cut; this select class is compatible with the discretization scheme defined by MERA. From this point of view, MERA defines a scheme for realizing discrete local conformal transformations in a theory with a cutoff. One additional caveat is that because MERA represents the wave-function at a given time, cuts through MERA can only implement `diagonal' conformal transformations, which preserve an equal time slice of the CFT.

It is useful to examine our results in reference to the TNR algorithm \cite{tnr}, which transforms the discretized Euclidean path integral to a MERA network \cite{tnr2mera}. Discretizing the path integral explicitly breaks conformal symmetry; if this symmetry is indeed represented by different cuts in MERA, it must be restored through the application of TNR. This is consistent with \cite{TNRlocal}, where the use of TNR to map the plane to a cylinder was seen to reproduce other features of local scale invariance, namely the correct spectrum of scaling dimensions from a transfer matrix in scale. Thus, our results add to the evidence in \cite{TNRlocal} that TNR restores the conformal invariance of the Euclidean path integral broken by the choice of discretization.

Our MERA representation of the thermal state on a circle is not the first; another one appeared in \cite{tnr2mera}. Let us briefly contrast the two constructions, leaving a more detailed comparison to Appendix~C. The quotient construction of the present paper proceeded in the following sequence: the path integral that prepares the ground state was discretized, turned into MERA via TNR \cite{tnr, tnr2mera}, passed through a logarithmic conformal map (eqs.~\ref{log2}-\ref{4log2} and figs.~\ref{fig:local}(b-c)), and finally quotiented. Ref.~\cite{tnr2mera} proceeded in a different order: the continuum path integral was first mapped logarithmically to a strip and only then discretized, with the subsequent steps (passing to the MERA description via TNR and taking the quotient) applied in either sequence. The equivalence of the two constructions tells us that conformal transformations commute with discretization---with the understanding that TNR and the procedure of adding or erasing tensors adopted in this paper implement conformal maps in the discretuum. 

The construction of the thermal state as a quotient of the vacuum is well known in holographic duality \cite{adscft}, where the gravitational dual of the thermal state on a circle (the BTZ black hole) is a geometric quotient of the dual to the vacuum (pure AdS$_3$ space) \cite{btzquotient}. Therefore, the quotient of the MERA network can be directly translated into the language of geometry. According to \cite{kinematicsp}, the three constituents of the quotient MERA -- the timelike, lightlike and spacelike-identified regions -- correspond to the geodesics that cross the black hole horizon, skirt it closely, and remain away from it. Of particular interest are the nearly isometric, lightlike-identified regions. Tensors that live in these regions see more than a full fundamental domain of open indices in their causal cone, so the coarse-graining they perform acts on scales larger than the spatial circle. The relevance of such super-IR scales for understanding black hole spacetimes was conjectured in \cite{entwinement} based on purely geometric considerations. The lightlike-identified regions of the quotient MERA are a tangible realization of this conjecture. The holographic interpretation of the quotient MERA will be discussed in more detail in a separate publication \cite{secondpaper}.

\textit{Acknowledgements.---}
We thank Lenny Susskind for useful discussions. BC, SM and JS thank Caltech, BC thanks the organizers of ``Holographic duality for condensed matter physics'' and KITPC-CAS in Beijing, and GE and GV thank Stanford University for hospitality. BC, GE, SM, XLQ, JS and GV thank the organizers of the programs ``Entanglement for Strongly Correlated Quantum Matter,'' ``Quantum Gravity Foundations: UV to IR'' and ``Closing the Entanglement Gap: Quantum Information, Quantum Matter and Quantum Fields'' held at KITP (supported in part by the National Science Foundation under Grant No. NSF PHY11-25915); BC, GE, LL, SM, XLQ, and JS thank the organizers of the ``Quantum Information Theory in Quantum Gravity II'' meeting held at the Perimeter Institute for Theoretical Physics; and 
BC, LL, SM and JS thank the organizers of the workshop ``AdS/CFT and Quantum Gravity'' at Centre de Recherches Math{\'e}matiques at the University of Montreal, where this work was carried out. SM was supported in part by an award from the Department of Energy (DOE) Office of Science Graduate Fellowship Program.
XLQ is supported by David and Lucile Packard Foundation. 
GE and GV acknowledge support by the Simons Foundation (Many Electron Collaboration). 
GV acknowledges support by the John Templeton Foundation and thanks the Australian Research Council Centre of Excellence for Engineered Quantum Systems. Research at Perimeter Institute is supported by the Government of Canada through Industry Canada and by the Province of Ontario through the Ministry of Research and Innovation.

\newpage

\section{Appendix A: Global and local scale transformations on the lattice}

\begin{figure}[!t]
\begin{center}
\includegraphics[trim = 4mm 3mm 4mm 3mm, clip, width=8.5cm]{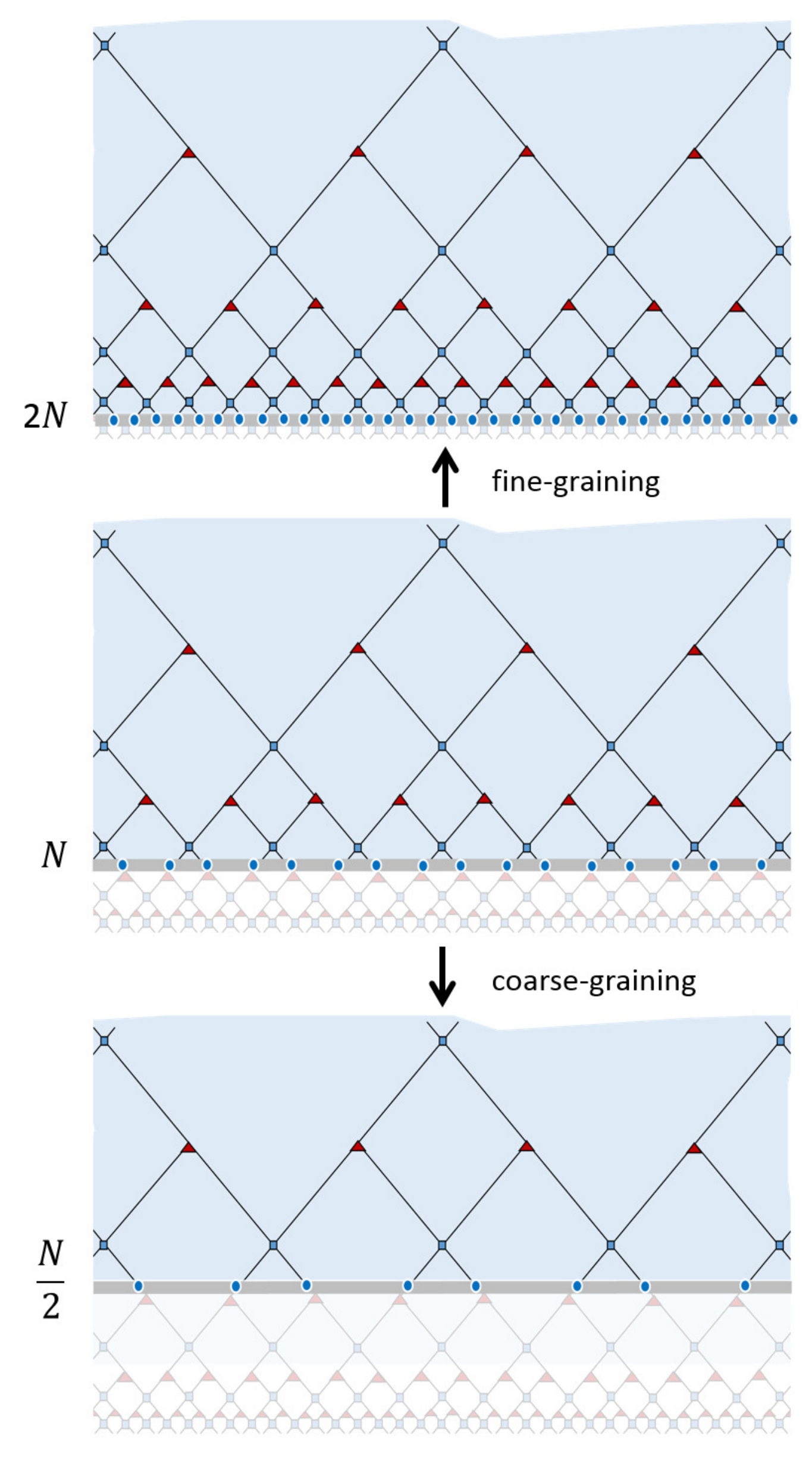}
\caption{
By adding or subtracting layers of the MERA, we can fine-grain or coarse-grain the lattice, thus implementing a global scale transformation by a discrete factor $1/2$ or $2$, respectively.}
\label{fig:global}
\end{center}
\end{figure}

MERA is a tensor network representation of many-body wave-functions that, upon optimization, approximates ground states of local Hamiltonians. Fig.~\ref{fig:global} shows MERA networks for one-dimensional systems. The open indices of the networks correspond to sites of the lattice on which the many-body wave-function is defined.

\textit{Global scale transformations.---} Each layer of tensors in the MERA consists of a row of disentanglers and a row of isometries and defines a coarse-graining transformation of the lattice. This coarse-graining implements a global scale transformation, in that it produces a renormalization group flow (in the space of ground states and, when applied by conjugation, in the space of local Hamiltonians) with the expected structure of RG fixed points, both at criticality and off criticality \cite{meraCFT}. These fixed-points are characterized by an explicitly scale invariant wave-function.

As illustrated in fig.~\ref{fig:global}, a layer of the MERA can be used to map the wave-function on a lattice with $N$ sites into the wave-function of a finer lattice with $2N$ sites, as well as into the wave-function of a coarser lattice with $N/2$ sites. Accordingly, the MERA does not define a single wave-function, but a collection of wave-functions, one for each of the lattices in a sequence of increasingly fine-grained (or coarse-grained) lattices. These wave-functions are all mutually consistent in that for any pair of wave-functions, there exists (by construction) a uniform fine-graining or coarse-graining transformation that maps one wave-function to the other. Notice that we can order all the wave-functions in this collection from most fine-grained to most coarse-grained.

\textit{Local scale transformations.---} Individual disentanglers and isometries can also be used to coarse-grain a region of the lattice without coarse-graining the rest. We would like to think of such a transformation as a local scale transformation. In order to show that individual disentanglers and isometries properly implement a local scale transformation, we investigate whether they indeed act on the lattice consistently with what we expect of a local scale transformation in the continuum. In a quantum critical system, the ground state in the continuum is often the vacuum of a conformal field theory. Conformal invariance in 1+1 dimensions implies in particular that applying certain local scale transformation to the ground state on the infinite line should result in a thermal state on the infinite line (and on a finite circle after a quotient). By showing that the same is true with the MERA, this paper provides strong evidence that we can use the disentanglers and isometries to properly implement local scale transformations.

Fig.~\ref{fig:local2} shows an example of the result of applying a local scale transformation to a MERA by adding and/or deleting disentanglers and isometries. Accordingly, the MERA defines an even larger collection of wave-functions than mentioned above. Each wave-function in this collection corresponds to a different way in which we locally coarse-grain the most refined lattice under consideration. All these wave-functions are once again mutually consistent, since by construction there is a local scale transformation mapping any pair of such wave-functions. Notice that this time we can define a \textit{partial order} in the set of wave-functions, according to whether one wave-function is more coarse-grained than another. This order is only partial because the map connecting the two wave-functions may require both coarse-graining one region and fine-graining another region, in which case none of the wave-functions is coarser or finer than the other.

\begin{figure}[!t]
\begin{center}
\includegraphics[width=8.5cm]{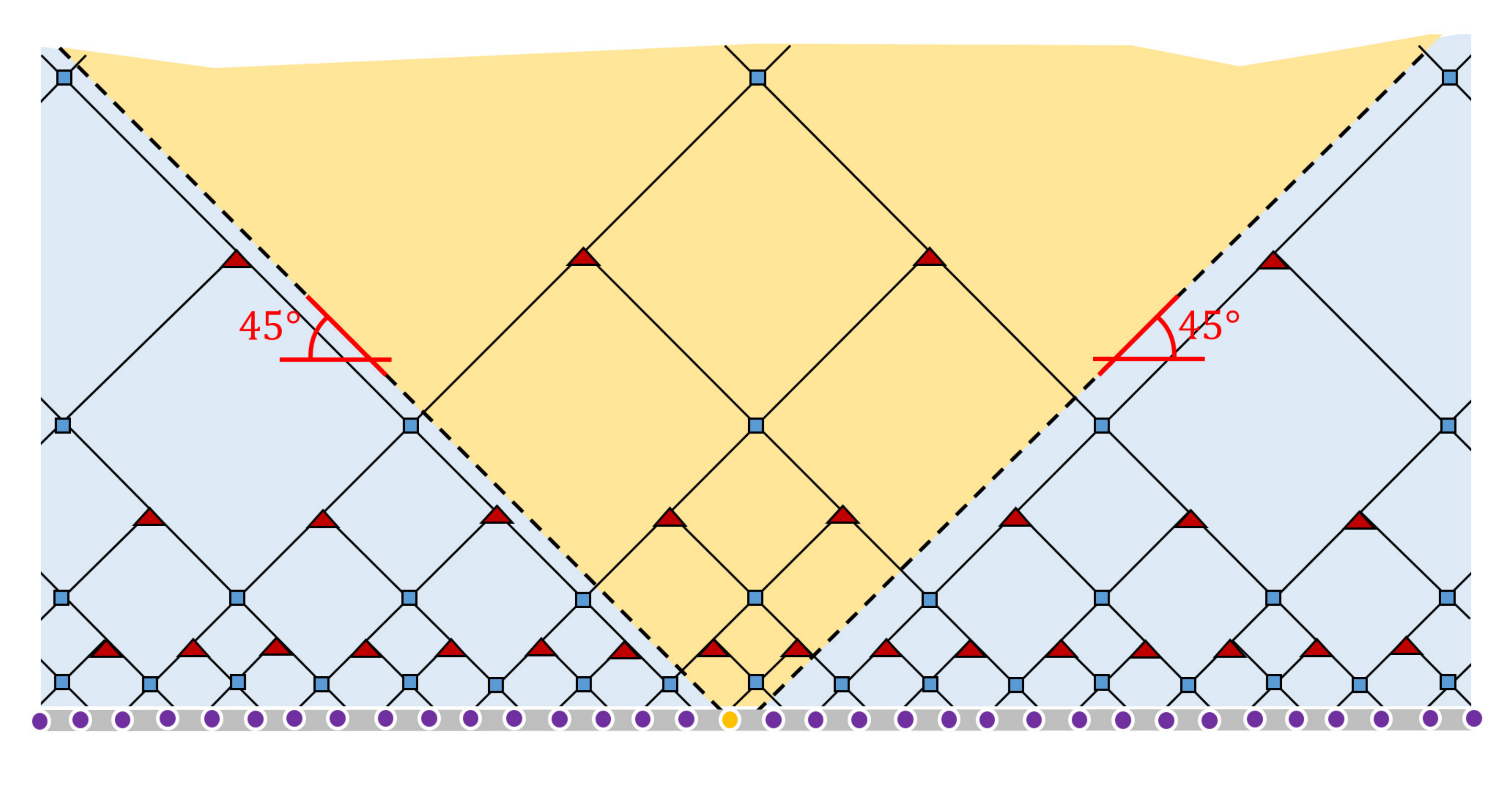}
\caption{
The causal cone of a site of the underlying lattice is the region of the tensor network that can affect the reduced density matrix on that site. Notice that the boundaries of the causal cone are at 45$^\circ$ (null or lightlike).
}
\label{fig:causal}
\end{center}
\end{figure}

\textit{Causal structure.---} The MERA can be understood as a quantum circuit  with some auxiliary time running from top to bottom \cite{mera}. The gates in this circuit are the disentanglers and isometries, which are unitary/isometric with respect to this auxiliary time. Given a site of the underlying lattice, we define its causal cone as the subset of gates in the quantum circuit that can influence the reduced density matrix on that site. Fig.~\ref{fig:causal} shows the causal cone of one site. More generally, we can define the quantum circuit of a set of sites of the lattice, which is seen to be the union of the causal cones for individual sites inside the region. Notice that the causal cone expands back in time at 45$^\circ$ with respect to the horizontal. We can think of 45$^\circ$ as corresponding to a null or lightlike direction, less than 45$^\circ$ as corresponding to a spacelike direction, and more than 45$^\circ$ as corresponding to a timelike direction. This analogy makes sense, since 45$^\circ$ defines the direction of propagation of information in the quantum circuit. 

It is easily seen that the MERA describes a wave-function for each lattice that can be obtained by a slice of the network that is piecewise lightlike and/or spacelike, but not timelike. Fig.~\ref{fig:local2} shows an example of a slice of the network that is piecewise horizontal and at 45$^\circ$.

\textit{Scale transformations and entanglement renormalization.---} We conclude this Appendix by pointing out that one can also coarse-grain a lattice (both globally and locally) by using instead the isometries of a \textit{tree tensor network} \cite{tree}-- which corresponds to a MERA with trivial disentanglers. In this case, however, at criticality we do not recover explicit invariance under global scale transformations or produce a thermal state under the local scale transformation studied in this paper. This can be traced back to the fact that a coarse-graining transformation based only on isometries fails to properly remove short-range entanglement. We thus conclude that the removal of short-range entanglement, as performed in the MERA by the disentanglers, is key to properly defining both global and local scale transformations on the lattice.

\begin{figure}[!t]
\begin{center}
\includegraphics[trim = 33mm 73mm 33mm 73mm, clip, width=8.5cm]{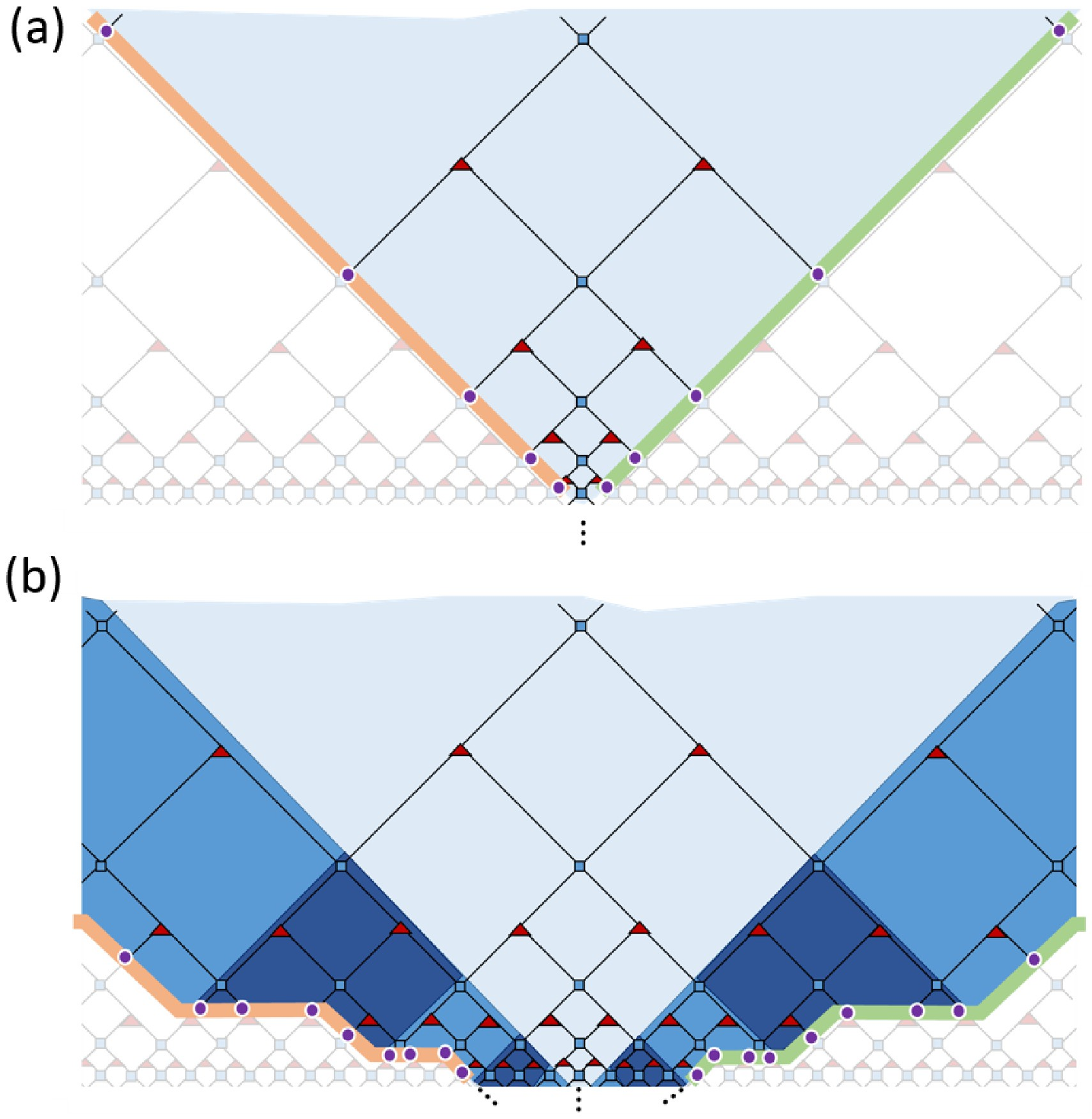}
\caption{
(a) Tensor network for the thermal state of inverse temperature $\beta = \pi/\log 2$.
(b) Tensor network for the thermal state of inverse temperature $4\beta$.
Notice that we can obtain one tensor network form the other by adding (or removing) a layer of tensors that consists of identical unit cells, each of which is made up of three disentanglers and three isometries and maps one site of (a) into four sites of (b).
}
\label{fig:explicit}
\end{center}
\end{figure}

\section{Appendix B: Exact scaling symmetry and quotient}

Under a discrete logarithmic scale transformation, the MERA representing the ground state on a discrete infinite line is taken to a MERA for a thermal state. This is illustrated in fig.~\ref{fig:explicit}. We re-display the two different implementations of the logarithmic transformation discussed in the main text, which correspond to inverse temperatures $\beta$ and $4\beta$, where $\beta = \pi/\log 2$. The figure emphasizes that the two resulting tensor networks are related by a relative global coarse-graining, formed by a uniform repetition of the same unit cell of tensors.

\begin{figure}[!t]
\begin{center}
\includegraphics[width=8.5cm]{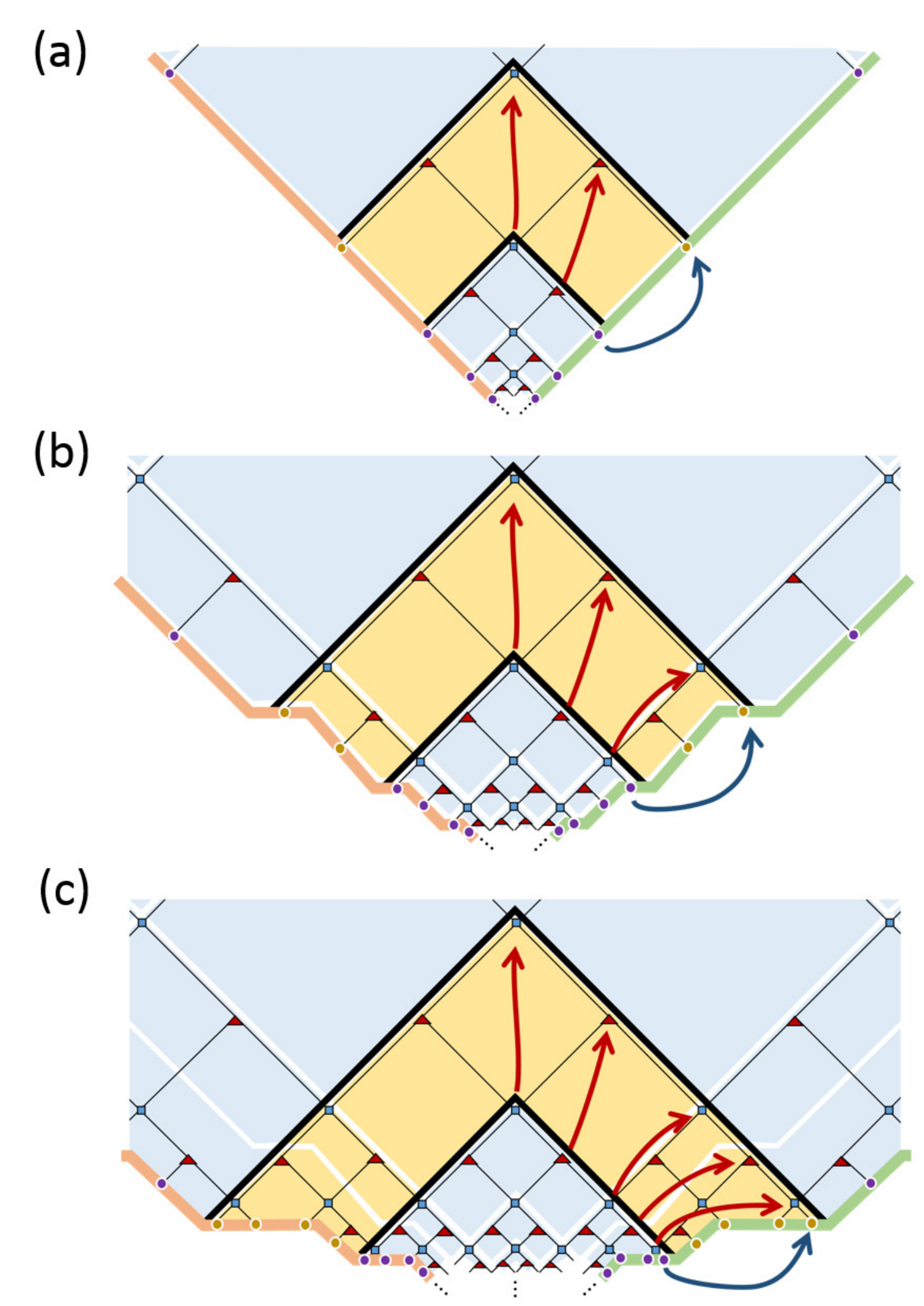}
\caption{
Tensor networks for a thermal state on the discrete infinite line, which result from applying a logarithmic local scale transformation to the MERA representing the ground state on the discrete infinite line. The networks in (a), (b) and (c) describe thermal states on a lattice that has a unit cell with $p$ sites, with $p=1,2,4$, respectively. As emphasized in fig.~\ref{fig:explicit}, these tensor networks are related: from network (b) we can produce network (a) or (c) by removing or adding a uniform row of tensors. By adding even more tensors, thermal states on a lattice with a larger unit cell can also be built. All these networks are invariant under translations by a unit cell.
}
\label{fig:symmetryx3}
\end{center}
\end{figure}

A symmetry of the tensor network for the thermal state with $\beta$, $2\beta$ and $4\beta$ is highlighted in fig.~\ref{fig:symmetryx3}. Each network is invariant under a finite scaling by a factor $1/2$ in the lattice where the MERA represents the ground state. In the logarithmic coordinate on the cut, this symmetry becomes a finite translation, as figs.~\ref{fig:commute} and \ref{fig:quotient1} of Appendix~C make explicit in the case of the $2\beta$ network. 
Notice that the MERA is in general an \textit{approximate} representation of the ground state on the lattice, one that can be made systematically more accurate by increasing the bond dimension of the tensors. However, the above symmetry is not approximate, but an \textit{exact} symmetry of the tensor network. 

\begin{figure}[!t]
\begin{center}
\includegraphics[width=7.5cm]{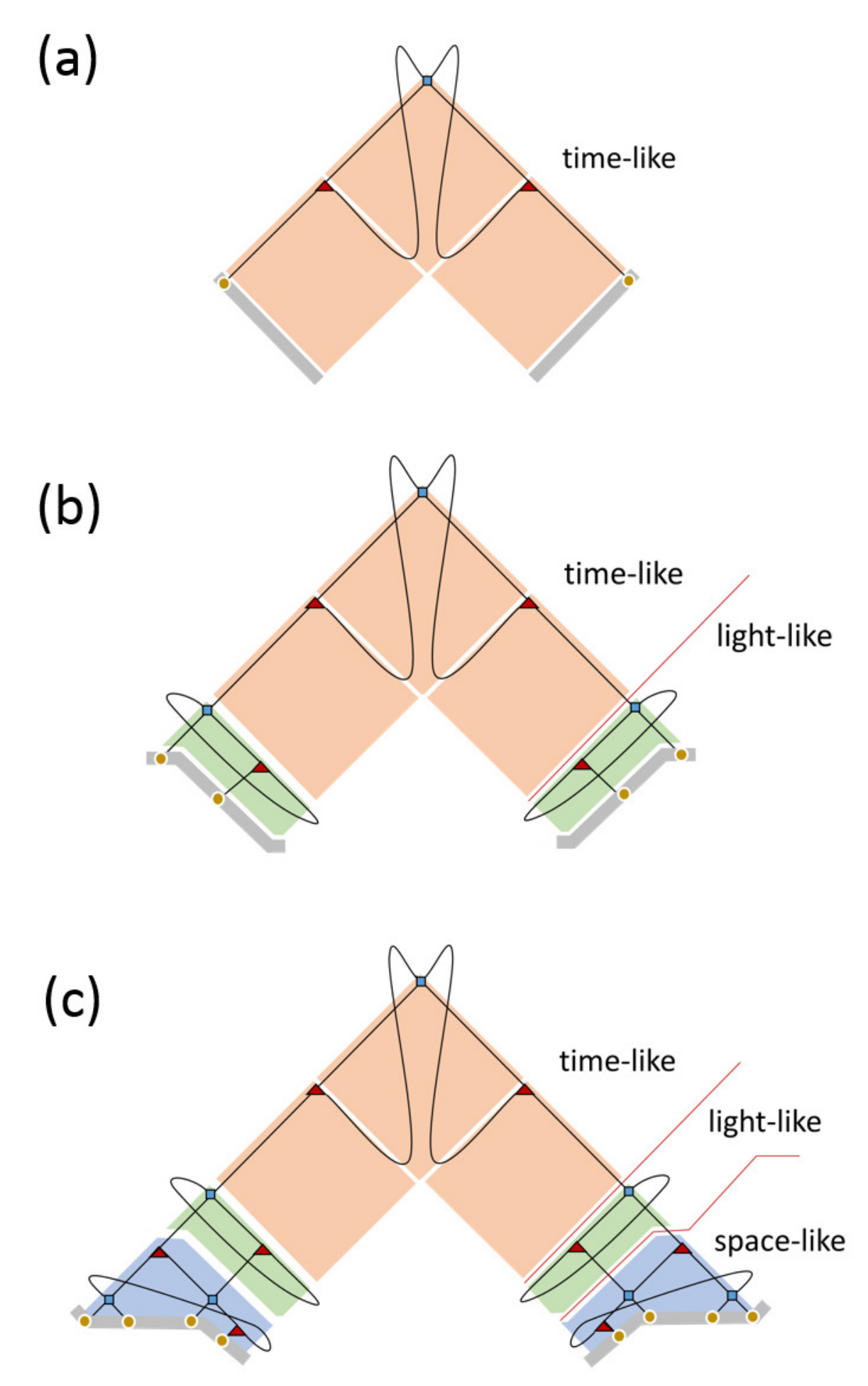}
\caption{
Quotient of the tensor networks in fig.~\ref{fig:symmetryx3} by a translation by a single unit cell ($k=1$), representing a thermal state on a lattice made up of $p$ sites, where $p=1,2,4$ for the networks (a), (b), and (c), respectively. 
}
\label{fig:quotientx3}
\end{center}
\end{figure}

Because this symmetry is exact, quotient by it by reconnecting lines. In the three networks of fig.~\ref{fig:symmetryx3}, the symmetry is a translation by $p$ sites, where respectively $p = 1, 2, 4$. We may quotient by a $k$-fold multiple of this translation. The result is a periodic tensor network describing a thermal state with the same inverse temperature (respectively, $\beta$, $2\beta$ and $4\beta$) but on a finite discrete circle made of $k\, p$ sites. Consequently, the reduced inverse temperature is independent of the choice of logarithmic map:
\begin{equation}
\tilde{\beta} = \frac{p\, \beta}{k\, p} = \frac{\pi}{k\log 2}.
\end{equation}
The tensor networks obtained from the quotient with $k=1$ and $p=1,2,4$ are shown in fig.~\ref{fig:quotientx3}. The quotient for $k=2$ is shown in fig.~\ref{fig:quotient}(b) of the main text.

\begin{figure}[!t]
\begin{center}
\includegraphics[width=8.5cm]{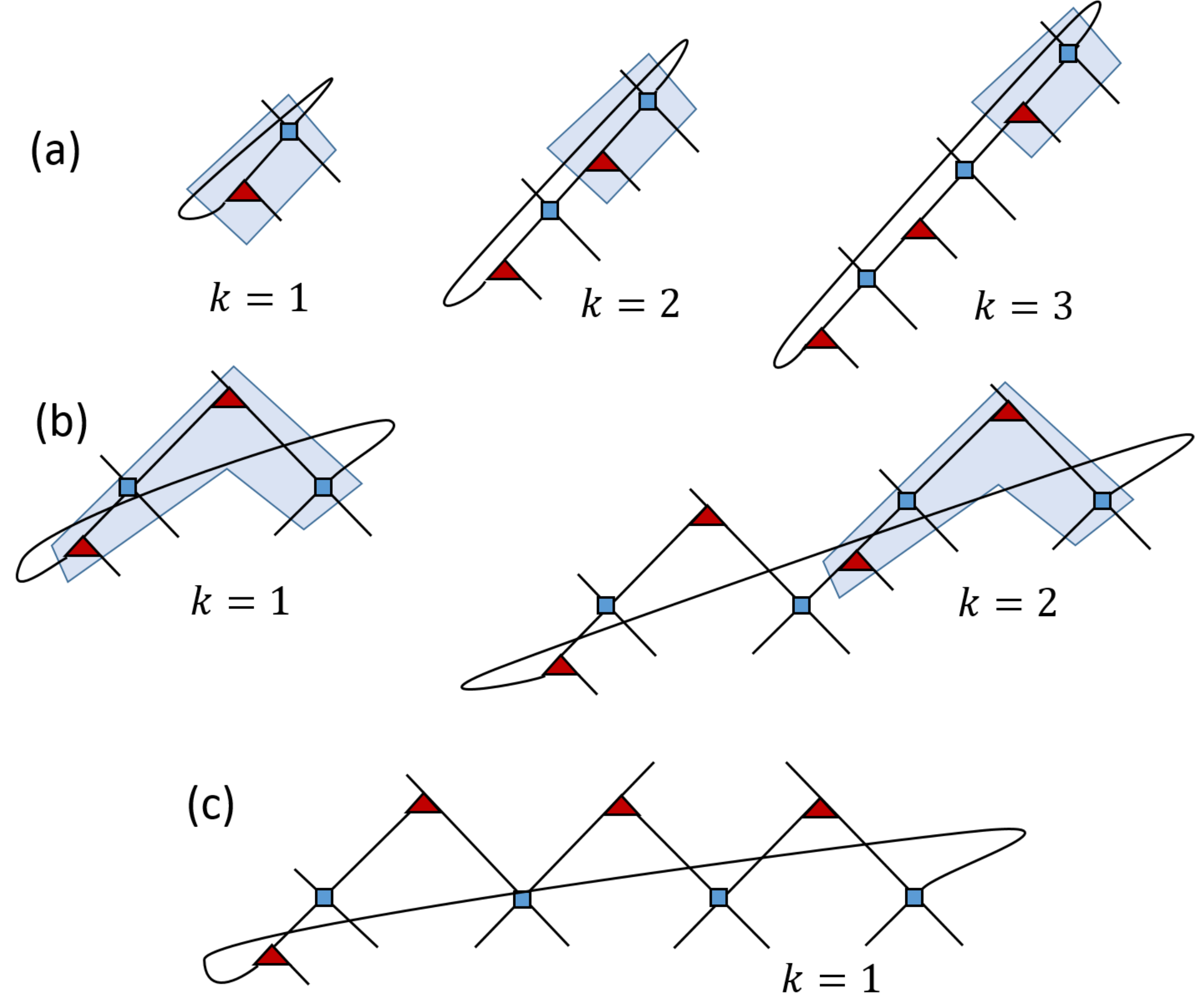}
\caption{
(a) Layer of tensors that map the lattice with a unit cell of $p=1$ sites into the lattice with a unit cell of $p=2$ sites, for $k=1$, $k=2$, and $k=3$. Notice that this part of the network is lightlike (45$^\circ$), with the periodic boundary conditions connecting the past and the future of a light ray.
(b) Layer of tensors that map the lattice with a unit cell of $p=2$ sites into the lattice with a unit cell of $p=4$ sites, for $k=1$ and $k=2$. This layer of tensors is an isometry, see fig.~\ref{fig:isometry}.
(c) Layer of tensors that map the lattice with a unit cell of $p=4$ sites into the lattice with a unit cell of $p=8$ sites, for $k=1$.}
\label{fig:layers}
\end{center}
\end{figure}

\begin{figure}[!t]
\begin{center}
\includegraphics[width=8.3cm]{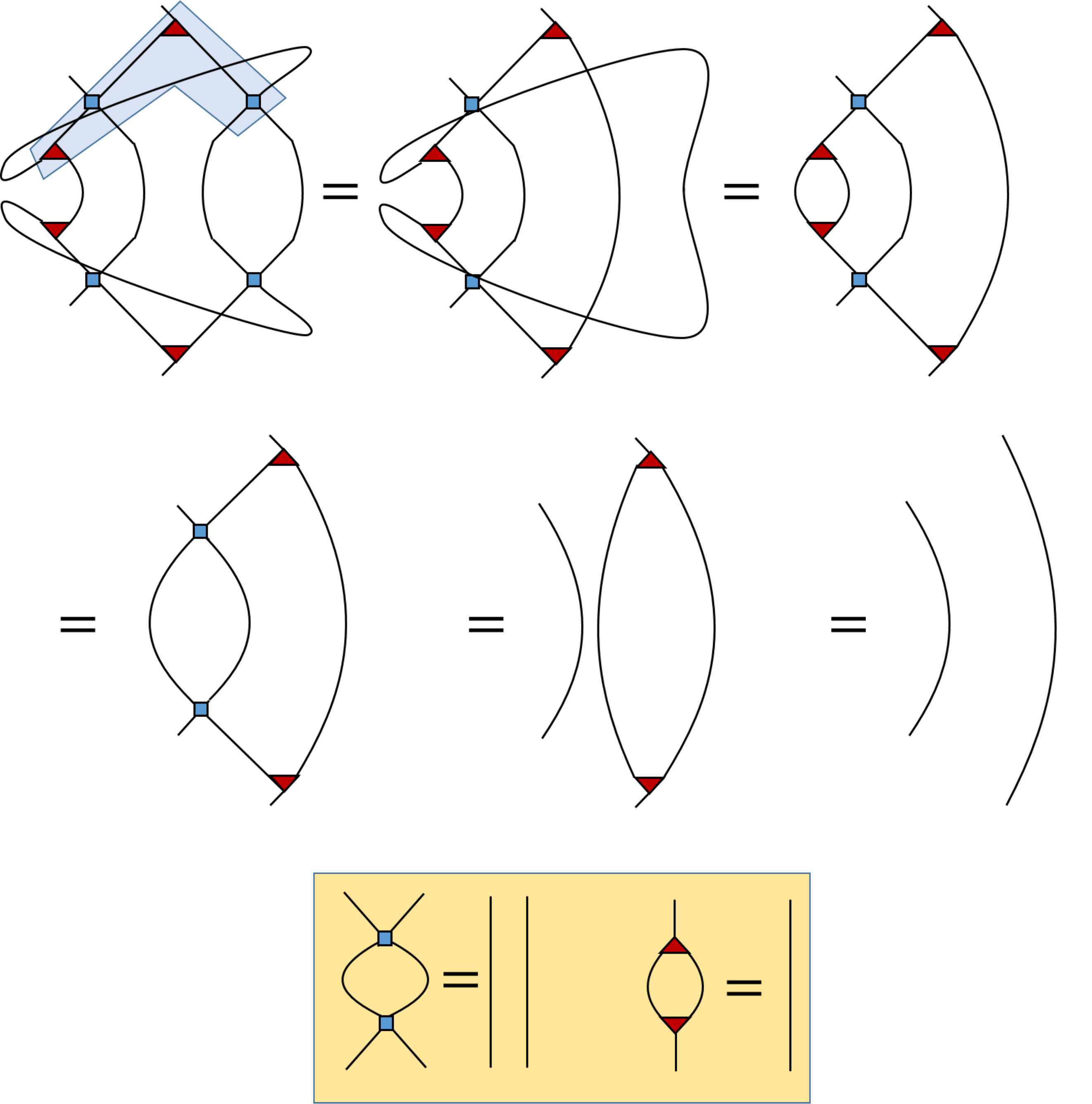}
\caption{
Sequence of replacements, which shows that the layer of disentanglers and isometries in fig.~\ref{fig:layers}(b left) is an exact isometry. Only the two equalities in the inset are used.
}
\label{fig:isometry}
\end{center}
\end{figure}

The quotient tensor network can be divided into three parts: a central core where the quotient reconnects tensors that are timelike-separated, a layer of lightlike-identified tensors, and a remainder, where the quotient acts spacelike. This last part consists of tensors organized in layers that implement an isometric fine-graining transformation; see fig.~\ref{fig:quotientx3}. Finally, fig.~\ref{fig:layers} shows more details of some of these layers of tensors while fig.~\ref{fig:isometry} shows with a concrete example that the spacelike layers are exact isometries.

\section{Appendix C: MERA quotient versus thermal MERA}
 
\begin{figure}[!t]
\begin{center}
\includegraphics[width=8.5cm]{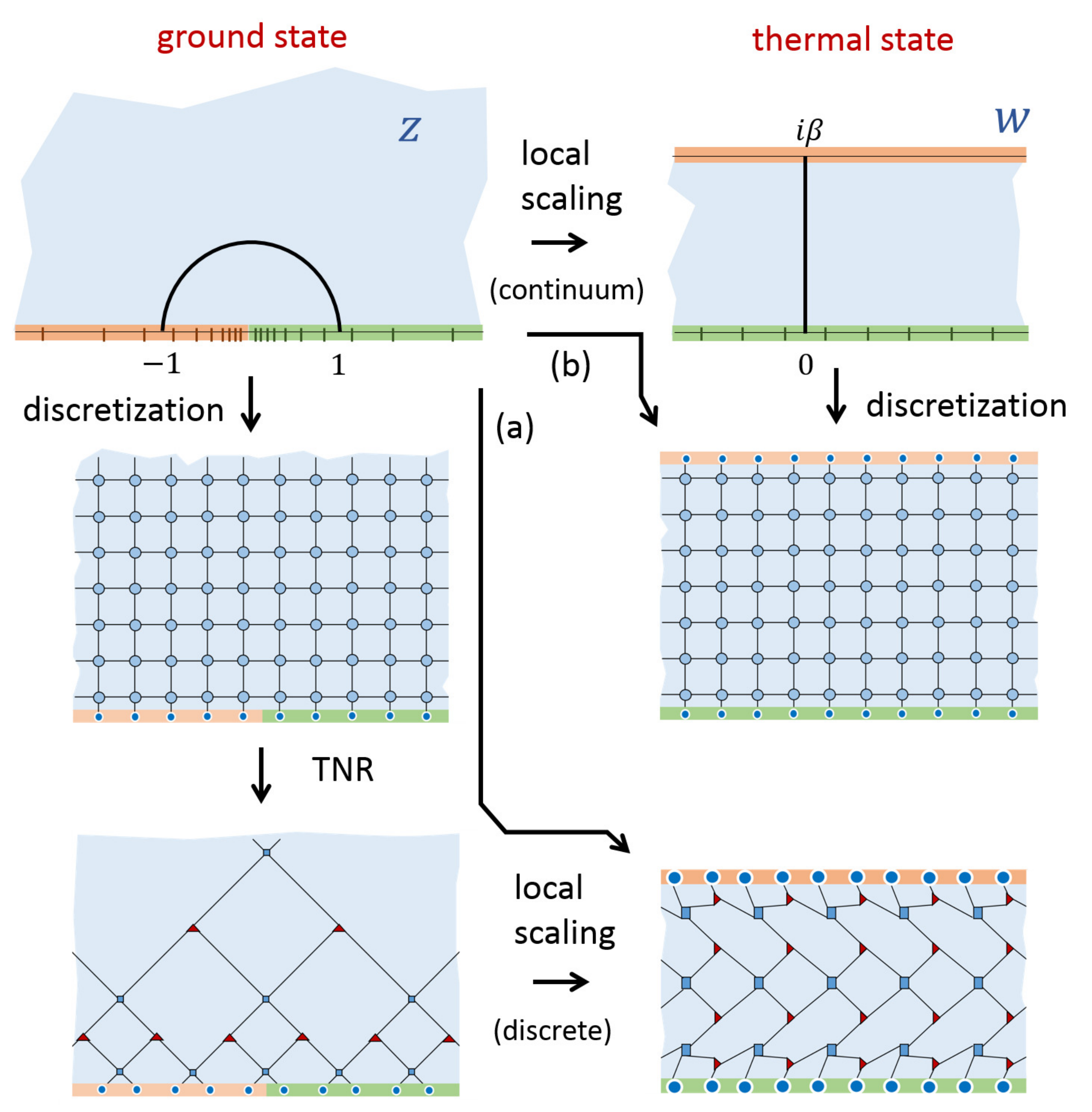}
\caption{
The Euclidean path integral on the upper half plane prepares the ground state on the infinite line. We can prepare a thermal state on the discretized infinite line in two ways: (a) By discretizing the Euclidean path integral on the upper half plane, then using TNR to produce a MERA for the ground state, then applying the logarithmic scaling transformation described in this paper. (b) By applying a logarithmic scaling transformation in the continuum to obtain the infinite strip and then discretizing the path integral.
}
\label{fig:commute}
\end{center}
\end{figure}

Here we contrast two different procedures for preparing a thermal state on the infinite, discretized line shown in figs.~\ref{fig:commute}(a) and (b). 

In the procedure outlined in fig.~\ref{fig:commute}(a), we start by discretizing the Euclidean path integral (on the upper half plane) that prepares the ground state of the continuous theory on the infinite line. This yields a discrete path integral (in the form of a square tensor network on the upper half plane) that prepares the ground state of the theory on an infinite one-dimensional lattice. Then we use TNR to transform the discrete Euclidean path integral into a MERA. The next step is the focus of the present paper. We apply an inhomogeneous change of cutoff by adding and removing tensors. This implements a lattice version of the conformal transformation 
$z \rightarrow (\beta/\pi) \log z$.

In the second procedure, shown in \ref{fig:commute}(b), we start by transforming the Euclidean path integral in the continuum with the conformal map $z \rightarrow (\beta/\pi) \log z$. This produces a Euclidean path integral on an infinite strip of width $\beta$, which prepares the thermal state of inverse temperature $\beta$ on the infinite line. After that, we discretize the Euclidean path integral by writing it as a square tensor network made of $\infty \times m_y$ tensors for some positive integer $m_y$.  Each tensor corresponds to an interval $\delta_x$ on the horizontal axis and $\delta_y \equiv \beta/m_y$ on the vertical axis.

\begin{figure}[!t]
\includegraphics[width=8.5cm]{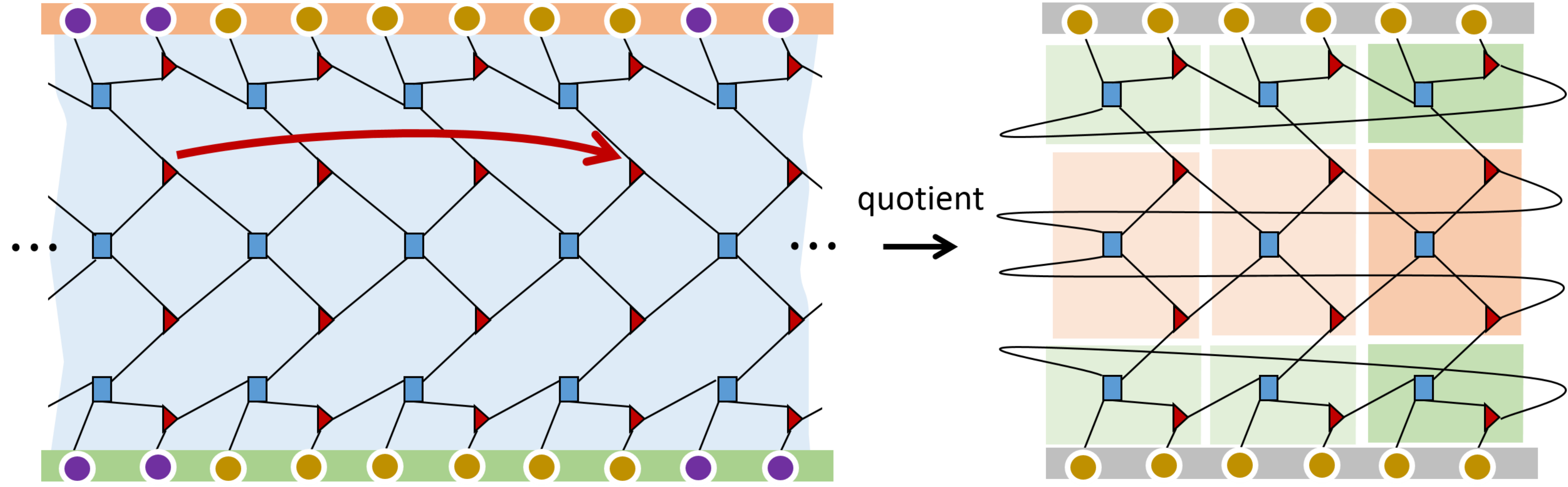}
\caption{
The transformed MERA tensor network on an infinite strip obtained through the procedure of fig.~\ref{fig:commute}(a). It represents a thermal state on the discretized infinite line. Because it is invariant under discrete translations, we may take its quotient to produce a tensor network for a thermal state on a spatial circle. 
}
\label{fig:quotient1}
\includegraphics[width=8.5cm]{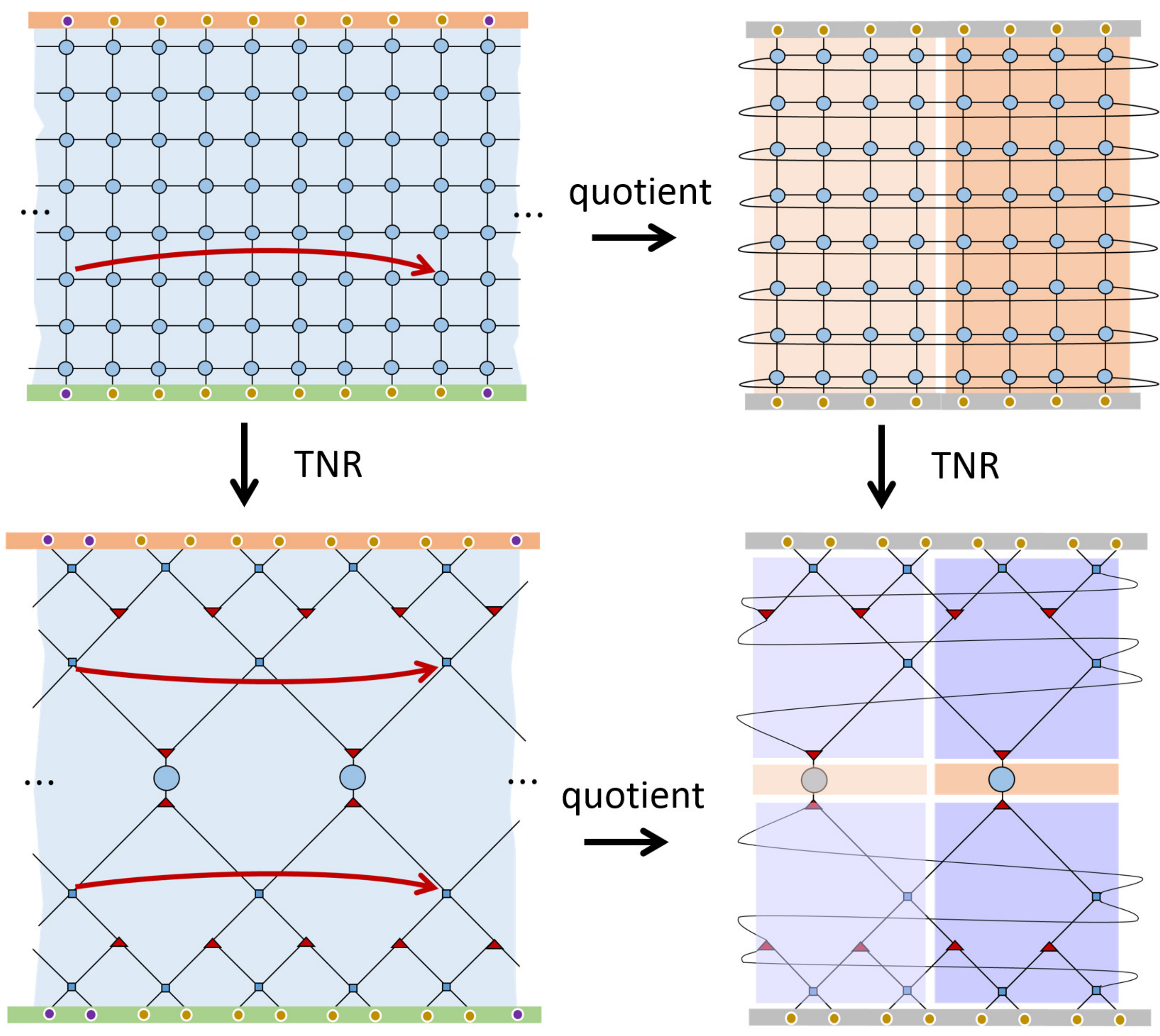}
\caption{
The discretized Euclidean partition function on an infinite strip obtained through the procedure of fig.~\ref{fig:commute}(b), which also represents a thermal state on the discretized infinite line. Ref.~\cite{tnr2mera} showed how to produce a MERA for this thermal state by applying TNR. The resulting thermal MERA is invariant under discrete translations, which allow us to take the quotient, producing a thermal state on a finite geometry. Alternatively, we can arrive to the same construction by first taking the quotient on the initial tensor network and only then using TNR to produce a MERA.
}
\label{fig:quotient2}
\end{figure}

Both procedures prepare a thermal state on the infinite, discretized line. In addition, we may now prepare a thermal state on a finite, discretized circle by modding out a discrete translation. This step is applied to the networks obtained from the two procedures in figs.~\ref{fig:quotient1} and \ref{fig:quotient2}, respectively.

The quotients may be taken, because both networks have an exact symmetry. The network produced in the first procedure is invariant under translations by a discrete amount $k \cdot \beta \log 2/\pi$, which corresponds to discrete scalings of the ground state by  $2^{k} = \exp(k\log 2)$ (for any positive integer $k$). Modding out a $k \cdot \beta \log 2/\pi$ translation as in fig.~\ref{fig:quotient1} produces a thermal state of inverse temperature $\beta$ on a circle of length $2\pi L = k \beta \log 2 /\pi $. Its reduced inverse temperature
\begin{equation}
\tilde{\beta} \equiv \frac{\beta}{2\pi L} = \frac{\pi }{k\log 2}
\end{equation}
determines the spectrum of the state according to eq.~(\ref{eq:spectrum}).

On the other hand, at the end of the second procedure the discretized Euclidean path integral is explicitly invariant under translations by $m_x$ tensors (or length $m_x \delta_x$) in the horizontal direction, for any positive integer $m_x$. Taking the quotient by a discrete $m_x \delta_x$ translation as in fig.~\ref{fig:quotient2} produces a network made up of $m_x \times m_y$ tensors, which represents a thermal state on a circle of length $m_x \delta_x$ and inverse temperature $\beta$. The reduced inverse temperature is $\tilde{\beta} = \beta / m_x\delta_x$. If we choose $m_x=k m_y$ for a positive integer $k$, then we have a rectangular tensor network for a thermal state with reduced inverse temperature $\tilde{\beta} = \delta_y/k \delta_x$. If, in addition, $m_y=2^q$ for some positive integer $q$, then we can apply $q$ rounds of TNR coarse-graining transformations to obtain a tensor network consisting of $q$ layers of MERA, a central row of $k$ coarse-grained tensors, and $q$ conjugated layers of MERA \cite{tnr2mera}.

With the choice $m_x \delta_x = k \cdot \beta \log 2 /\pi$, both constructions produce the thermal state with inverse temperature $\beta$ on a finite circle of length $k \cdot \beta \log 2 /\pi$ whose reduced inverse temperature is $\tilde{\beta}= \pi / k \log 2$.

\end{document}